# Novel drying additives and their evaluation for self-flowing refractory castables

B. P. Bezerra[1,*]; A. P. Luz[1], V. C. Pandolfelli[1]

[1]Federal University of Sao Carlos, Graduate Program in Materials Science and Engineering

Rod. Washington Luis, km 235 - São Carlos - SP - Brazil - CEP:13565-905

*Corresponding author at: t*el.:* +55-16-33518601

E-mail: bezerrap.breno@gmail.com

## Abstract

The drying step of dense refractory castables containing hydraulic binders is a critical process, which usually requires using slow heating rates due to the high explosion trend of such materials during their first thermal treatment. Thus, this work investigated the performance of alternative additives to induce faster and safer drying of self-flowing high-alumina refractory castables bonded with calcium aluminate cement (CAC) or hydratable alumina (HA). The following materials were analyzed for this purpose: polymeric fibers, a permeability enhancing compound (RefPac MIPORE 20) and an organic additive (aluminum salt of 2-hydroxypropanoic acid). The drying behavior and explosion resistance of the cured samples were evaluated when subjecting the prepared castables to heating rates of 2, 5 or 20°C/min and the obtained data were then correlated to the potential of the drying agents to improve the permeability and mechanical strength level of the refractories at different temperatures. The collected results attested that the selected additives were more efficient in optimizing the drying behavior of the CAC-bonded compositions, whereas the HA-containing castables performed better when the aluminum-based salt was blended with a small amount of CAC (0.5 wt.%), which changed the binders hydration reaction sequence and optimized the permeability level of the resulting microstructure. Consequently, some of the designed compositions evaluated in this work showed improved drying behavior and no explosion was observed even during the tests carried out under a high heating rate (20°C/min).

# 1. Introduction

The growing demand for advanced refractory castables in recent decades has been motivated by the requirements of industrial processes operating at high temperatures, such as iron and steelmaking, cement and glasses production, petrochemical, and others, which has helped to develop new technologies and innovations in the design of high-performance ceramic systems.

Although the continuous progress of refractory castables has focused on finding alternatives to reduce the use of hydraulic binders in formulations, calcium aluminate cement (CAC) is still the most used one in a large number of commercial products due to its ability to provide higher green mechanical strength, presenting a suitable setting time (around 6-24h) and chemical resistance to a variety of corrosive environments [1-3]. Hydratable alumina (HA) is another binder usually applied in castable compositions, as it helps to design products with high refractoriness, without the drawbacks (i.e., generation of phases with low melting point) associated with the presence of CaO, derived from CAC, in formulations containing alumina and silica [4-6]. During HA hydration, a thick layer of amorphous gel is initially generated on the alumina's surface and it may be partially crystallized at higher temperatures or over a longer time, into boehmite [$Al_2O_3.(1-2)H_2O$] and bayerite ($Al_2O_3.3H_2O$). Besides that, the gel phase is able to fill in the pores and defects of the microstructure [3,7], reducing the refractory's permeability and, consequently, its explosion resistance during the drying step.

As a consequence of the low permeability of dense castables bonded with hydraulic binders, spallings or even explosions could take place during the first heating treatment of these ceramics, depending on the selected heating schedule. Such behavior is related to the steam pressure derived from the free-water release and decomposition of hydrated phases at temperatures above 100°C, as the entrapment of the gas phase in the resulting microstructure during heating may easily lead to pressure levels that exceed the green mechanical strength of the ceramic. Thus, in order to avoid this undesired condition, the drying schedule of dense refractory castables are usually based on various dwell steps at intermediate temperatures in the 100-600°C range, which leads to very conservative heating procedures.

Aiming to optimize the energy efficiency and to reduce the cost associated with this first thermal treatment of the refractories, various attempts have been carried out in order to adjust the permeability of such materials and provide easier and safer vapor release. Some alternatives proposed in the literature are using additives that may act generating permeable channels in the consolidated microstructure, favoring the percolation of the pressurized fluids during the first heating step of these materials. Among the proposed additives, using polymeric fibers with a low melting point and decomposition temperature is the most common and well-known solution presented in the literature [8-10]. For instance, polyethylene [11] and polypropylene fibers are likely drying agents for castables due to their low melting point (100-120°C and > 150°C, respectively).

Additionally, a new sort of additive for enhancing the drying ability of CAC-bonded castables has been reported in the literature, called RefPac Mipore 20 (MP) [12,13]. This product acts by modifying the cement hydration reaction sequence, giving rise to gel-like phases that decompose mainly around 100-150°C, inducing the formation of a more permeable microstructure that allows the steam release from the inner region up to the material's surface, during the initial stages of the heating process [14]. The role of this additive in castables containing other hydraulic binders is still not well described, although Luz *et al.* [15] evaluated the incorporation of MP into vibratable high-alumina HA-bonded refractories.

Aluminum salt 2-hydroxypropanoic acid [$Al(CH_3CHOHCOO)_3$] is also an interesting anti-explosion agent, which can increase the number of permeable paths in magnesia-based refractory products [16-18]. However, the acting mechanism of this compound is not clearly reported and there is a lack of studies using this additive in compositions containing hydraulic binders.

Based on these aspects, this work investigates: (*i*) the influence of two novel commercial drying agents commercially available [polyethylene fibers (Emsil-Dry) and the active compound called RefPac MiPore 20] in the drying performance of self-flowing high-alumina refractory castables (containing calcium aluminate cement or hydratable alumina as binders); and (*ii*) the feasibility of incorporating aluminum salt 2-hydroxypropanoic acid [$Al(CH_3CHOHCOO)_3$] into hydratable alumina-bonded castables, as an alternative route to obtain more permeable microstructures, which could be less prone to undergo explosion under faster heating schedules.

## 2. Experimental procedure

### 2.1 Raw materials selection and designed compositions

High-alumina refractory castables were designed based on the Andreasen's particle packing model [19,20], as described in Eq. 1, and considering the distribution coefficient (*q*) equal to 0.21, in order to obtain self-flowing compositions.

$$\frac{CPFT}{100} = \left(\frac{D}{D_L}\right)^q \quad (1)$$

where *CPFT* is the cumulative percentage of particles finer than *D*; *q* is the distribution coefficient; *D* is the particle size; and $D_L$ is the largest particle size.

The selected raw materials for the studied formulations consisted of tabular alumina ($D \leq 6$ mm, Almatis, Germany), calcined and reactive aluminas (CL370 and CT3000SG, respectively, supplied by Almatis, Germany), as indicated in Table 1. Calcium aluminate cement (CAC, Secar 71, Imerys Aluminates, France) or hydratable alumina (HA, Alphabond 300, Almatis, Germany) were the binders evaluated in this work. In order to improve the drying ability of the castables, 0.1 wt.% of polyethylene fibers with length < 6 mm (Emsil Dry, Elkem, Norway) or 2.5 wt.% of the active compound Refpac Mipore 20 (MP, Imerys Aluminate, France) were incorporated into the CAC-bonded compositions. The latter is a powdered additive ($d_{50}$ ~10-20 μm), comprised by organic compounds mixed with mineral phases containing $Al_2O_3$ (39-43 wt.%), CaO (12-15 wt.%) and MgO (16-20 wt.%) [12]. Previous investigations [13,14] reported that 2.5 wt.% of MP in replacing the CAC content is a suitable amount to provide improved flowability, green mechanical strength and high permeability for the castables. For this reason, the same content of MP was selected to be used in the designed compositions of this work.

Table 2 presents the compositions prepared for the evaluation of the aluminum salt [AS, $Al(CH_3CHOHCOO)_3$, Quimibras Industrias Químicas S.A., Brazil] as a drying agent for high-alumina castables bonded with hydratable alumina (HA). According to some investigations presented in the literature [16-18], 0.3-0.5 wt.% of AS may be sufficient to modify the permeability of MgO-bonded castable compositions. Hence, it was decided to analyze the addition of 0.3 or 0.6 wt.% of this organic compound to the designed alumina-based systems.

During the mixing step of the refractories, distinct water contents were incorporated into the mixes (Table 1 and 2), as well as 0.2 wt% of a polymeric dispersant (Castament® FS60, BASF, Germany). The only exception was related to the compositions containing RefPac MiPore 20, as this drying agent already contains dispersant compounds among its components [13].

Table 1 – General composition of the refractory castables containing polymeric fibers (Emsil-Dry) or MiPore (MP) as a drying agent.

| Raw materials | Compositions (wt.%) | | | | | |
| --- | --- | --- | --- | --- | --- | --- |
| | 5CAC | 5HA | 5CAC-ED | 5HA-ED | 2.5CAC-MP | 2.5HA-MP |
| Tabular alumina | 74 | 77 | 74 | 77 | 74 | 77 |
| Calcined alumina (CL370) | 11 | 8 | 11 | 8 | 11 | 8 |
| Reactive alumina (CT3000SG) | 10 | 10 | 10 | 10 | 10 | 10 |
| Calcium aluminate cement (Secar 71) | 5 | - | 5 | - | 2.5 | - |
| Hydratable alumina (Alphabond 300) | - | 5 | - | 5 | - | 2.5 |
| Emsil-Dry | - | - | 0.1 | 0.1 | - | - |
| RefPac MiPore 20 | - | - | - | - | 2.5 | 2.5 |
| *Distilled water* | 4.5 | 5.2 | 4.6 | 5.1 | 4.8 | 5.9 |
| *Free-flow (%)* | 89 | 85 | 81 | 86 | 102* | 108* |

*vibratable flow

Table 2 – General information of the castables bonded with hydratable alumina and containing the aluminum salt as a drying agent.

| Raw materials | Compositions (wt.%) | | | | |
| --- | --- | --- | --- | --- | --- |
| | 5HA | 5HA-0.3AS | 5HA-0.3AS-0.5CAC | 5HA-0.6AS | 5HA-0.6AS-0.5CAC |
| Tabular alumina | 77 | 77 | 77 | 77 | 77 |
| Calcined alumina (CL370) | 8 | 8 | 7.5 | 8 | 7.5 |
| Reactive alumina (CT3000SG) | 10 | 10 | 10 | 10 | 10 |
| Calcium aluminate cement (Secar 71) | - | - | 0.5 | - | 0.5 |
| Hydratable alumina (Alphabond 300) | 5 | 5 | 5 | 5 | 5 |
| Aluminum salt of 2-hydroxypropanoic acid (AS) | - | 0.3 | 0.3 | 0.6 | 0.6 |
| *Distilled water* | 5.2 | 5.7 | 5.7 | 5.7 | 5.7 |
| *Free-flow (%)* | 85 | 93 | 93 | 99 | 101 |

## 2.2 Refractory castables' processing and characterization

The compositions with and without additives were prepared in a planetary rheometer [21] and the mixing procedure consisted of dry and wet homogenization stages. After that, the curing behavior and setting time of the fresh castables' mixes were characterized by measuring the propagation velocity of ultrasonic waves (Ultrasonic Measuring Test System IP-8, Germany) in the compositions as a function of time at room temperature (~22°C). The prepared castables were also molded as: *(i)* cylinders (50 x 50 mm$^2$) for the explosion and drying tests; *(ii)* discs (d = 75 mm, h = 26 mm) for the permeability measurements; and *(iii)* bars (150 x 25 x 25 mm$^3$) for the 3-point bending tests and apparent porosity characterization.

The samples' curing was carried out at 30°C for 24h or 36h and the CAC-containing compositions were kept in closed plastic bags containing a beaker with water in order to maintain a humid environment. After that, the castables were demolded and dried at 110°C for 24h and some of the specimens were calcined in the 200-400°C range, using a heating rate of 1°C/min and dwell time of 5h at the selected temperature.

The drying behavior and explosion resistance of cured samples was carried out in a device that simultaneously measured the thermogravimetric profiles and the samples' temperature [22], when applying heating rates of 2, 5 and 20°C/min up to 600°C. On the other hand, permeability tests were conducted at room temperature [23] using samples obtained after drying at 110°C/ for 24h and calcining at 200, 300 and 450°C for 5h. The Darcian ($k_1$) and non-Darcian ($k_2$) permeability constants were calculated by fitting the $(P_e^2 - P_s^2)/(2P_s L)$ values as a function of the percolating fluid velocity ($v_s$) to a polynomial function, as described by the Forchheimer equation [24].

$$\frac{P_e^2 - P_s^2}{2P_s L} = \frac{\mu}{k_1} v_s + \frac{\rho}{k_2} v_s^2 \qquad (2)$$

where $P_e$ and $P_s$ are, respectively, the inlet and outlet pressures of the air-flow; $L$ is the samples' thickness, $\mu$ (1.8 x 10$^{-5}$ Pa.s) is the air viscosity and $\rho$ (1.08 kg/m$^3$) is its density at room temperature.

Additionally, cold modulus of rupture (3 point-bending tests, ASTM C133-97) and apparent porosity (ASTM C830-00) of the compositions containing polymeric fibers or Refpac Mipore 20 (Table 1) were determined after curing, drying and calcining (200, 300 and 400°C for 5h) these refractories. On the other hand, the formulations containing aluminum salt were subjected to the same experimental tests, but their properties were only evaluated after curing (30°C for 24 or 36h) and drying (110°C for 24) steps. A total of five samples for each composition and selected firing temperature were analyzed during these measurements.

# 3 Results and discussion

## 3.1 - Effect of the polymeric fibers (ED) and active compound (MP) in the castables' properties

The curing behavior and setting time of the prepared castables were analyzed by recording the ultrasonic wave propagation velocity results as a function of time at 22°C for 24h. As shown in Fig. 1a, the reference composition containing cement (5CAC) presented a great increase in the measured velocity between 5 to 13 hours, whereas the hardening process of the one bonded with hydratable alumina (5HA, Fig. 1b) began after 3 hours, nevertheless presenting lower velocity values in the evaluated period of time.

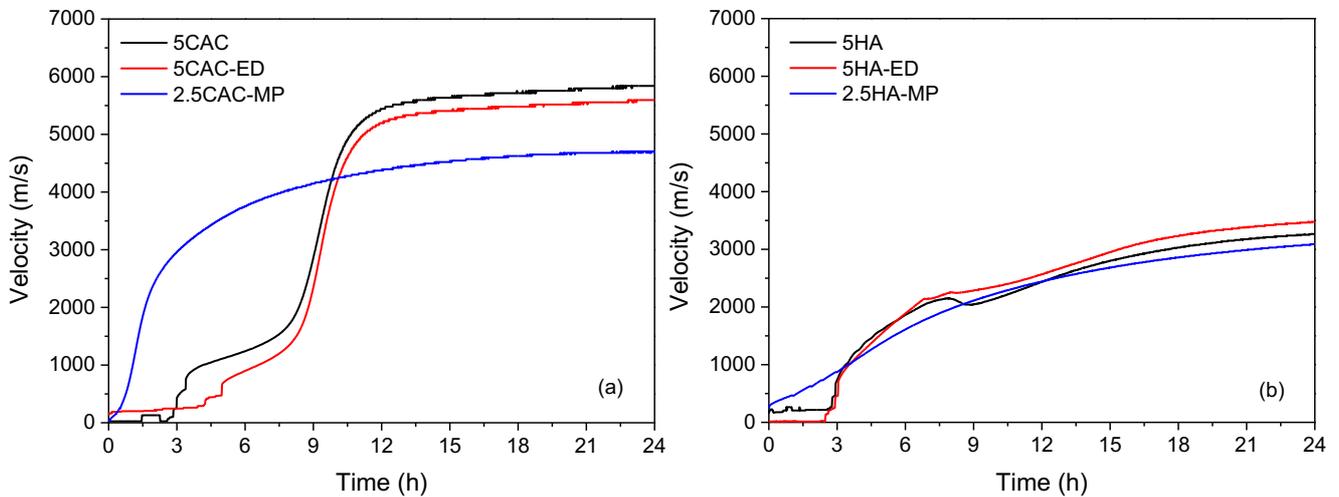

Figure 1 – Curing behavior of the designed high-alumina castables bonded with (a) calcium aluminate cement (CAC) or (b) hydratable alumina (HA). ED = polymeric fibers and MP = Refpac Mipore 20. The samples were maintained at 22°C for 24h.

This performance may be related to the differences in the hydration process of the tested hydraulic binders. For instance, CAC hydration usually gives rise to hydrated compounds (i.e., $CAH_{10}$, $C_2AH_8$, $C_3AH_6$ and/or $AH_3$, where C = CaO, A = $Al_2O_3$ and H = $H_2O$) and the likelihood of which one of them will be precipitated in the refractory microstructure is associated with the curing time and temperature conditions, as well as the water content used during the mixing stage and the humidity level available in the curing environment [12]. In order to favor the precipitation of the most stable hydrates (crystalline $C_3AH_6$ and $AH_3$), curing temperatures above 35°C are suggested, which should result in refractories with improved green mechanical strength [2]. As the ultrasonic measurements were carried out at room temperature (22°C), other metastable hydrated compounds (such as $CAH_{10}$, $C_2AH_8$ and gel-like $AH_3$) may also be generated in the designed compositions, and the precipitation of these compounds is responsible for the setting and hardening behavior of the CAC-bonded castables (Fig. 1a).

On the other hand, the hydration of hydratable alumina is based on the generation of gel-like hydrated phases, which can crystallize and mainly result in the formation of $Al_2O_3.3H_2O$ and $Al_2O_3.(1-2)H_2O$. Such transformations usually lead to microstructures with limited permeability level and lower green mechanical strength than the ones derived from CAC hydration process [7]. As confirmed in Fig. 1b, indeed HA-bonded castables presented lower velocity results than the compositions containing CAC, which can be related to the development of samples' with lower elastic modulus for the former formulations.

The incorporation of the polymeric fibers (ED) into the designed compositions did not induce significant changes in the curing behavior of the castables, as shown in Fig. 1. However, the presence of the MP additive modified the hydration reaction sequence of the binders [12-15], which affected the setting time of the castables, speeding up their hardening.

A more cohesive microstructure is associated to higher mechanical strength and lower permeability level, and both aspects have important roles on the drying behavior of ceramic linings. For instance, the HA-bonded refractories presented lower flexural strength (Fig. 2b) than the ones with CAC (Fig. 2a), which indicates that such materials are more prone to fail or explode, as they will be able to withstand limited steam pressure levels during heating. Moreover, higher water content was added to the hydratable alumina-containing compositions (Table 1), increasing the refractories' pressurization likelihood.

The mechanical strength of the analyzed castables increased when drying the samples at 110°C or firing them at different temperatures in the range of 200-400°C (Fig. 2), most likely due to the crystallization of the gel-like hydrates and formation of interconnected crystalline hydrated phases in the microstructure at higher temperatures, favoring binding among the adjacent coarse and fine components of the composition [3, 6]. Although the release of free-water and the decomposition of the hydrated compound may induce the samples' porosity increase, the binding effect enhanced the overall cold flexural strength values of the designed refractories.

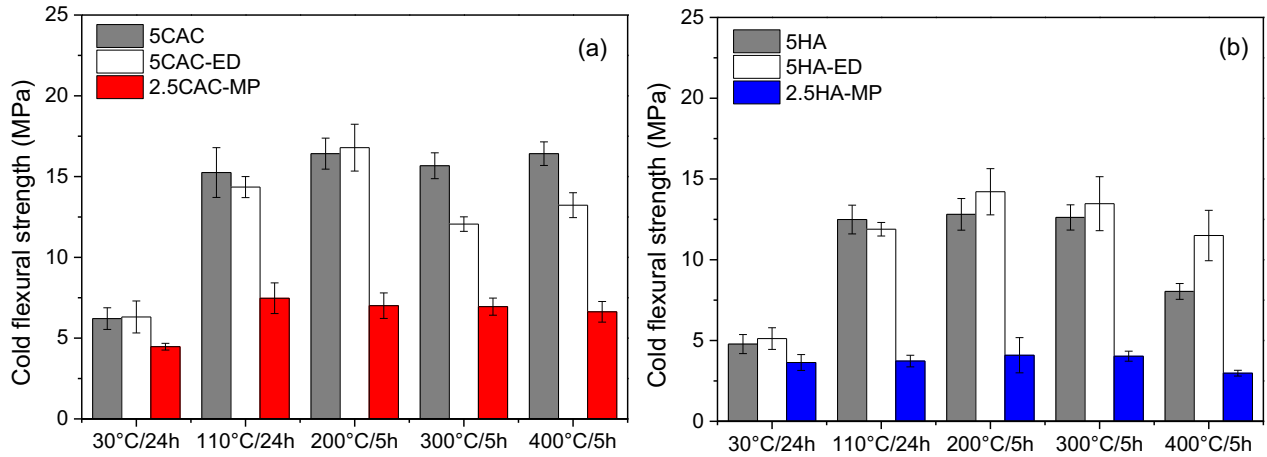

Figure 2 – Cold flexural strength of the evaluated castables obtained after curing (30°C/24h), drying (110°C/24h) and firing (200-400 °C/5h).

Regarding the additive-containing refractories, adding fibers to 5CAC-ED and 5HA-ED did not affect the mechanical strength of the samples to a major extent (Fig. 2). Nevertheless, the permeability enhancing compound (MP) led to the decrease in the castables' modulus of rupture in all tested conditions, when compared to the reference materials (5CAC or 5HA). Such behavior has already been reported in previous studies [12-15], and it was related to the interaction of this additive with the binders, which changes the hydration reaction sequence and the precipitated phases in the resulting microstructure.

Despite this side effect, MP is very effective in increasing the permeability level of the 2.5CAC-MP and 2.5HA-MP compositions, favoring the development of high $k_1$ and $k_2$ (permeability constants) values even for the materials just dried at 110°C for 24h or calcined up to 400°C for 5h (Fig. 3). On the other hand, the reference refractories (5CAC or 5HA) presented very low permeability in all evaluated conditions, which prevented the measurements of $k_1$ and $k_2$ for the 5HA samples after drying at 110°C and firing at 200°C for 5h, when using the chosen permeameter equipment [24]. Hence, in such cases (110°C and 200°C), the permeability constant values for 5HA were lower than the ones obtained for the same composition when fired at 300°C for 5h ($k_1$ = 8.98 x $10^{-18}$ $m^2$ and $k_2$ = 4.70 x $10^{-18}$ m).

In general, the incorporation of the polymeric fibers to 5CAC-ED and 5HA-ED compositions only provided a slight increase of the overall permeability constants, maintaining the measured values in the same order of magnitude as the reference materials (5CAC and 5HA, Fig. 3).

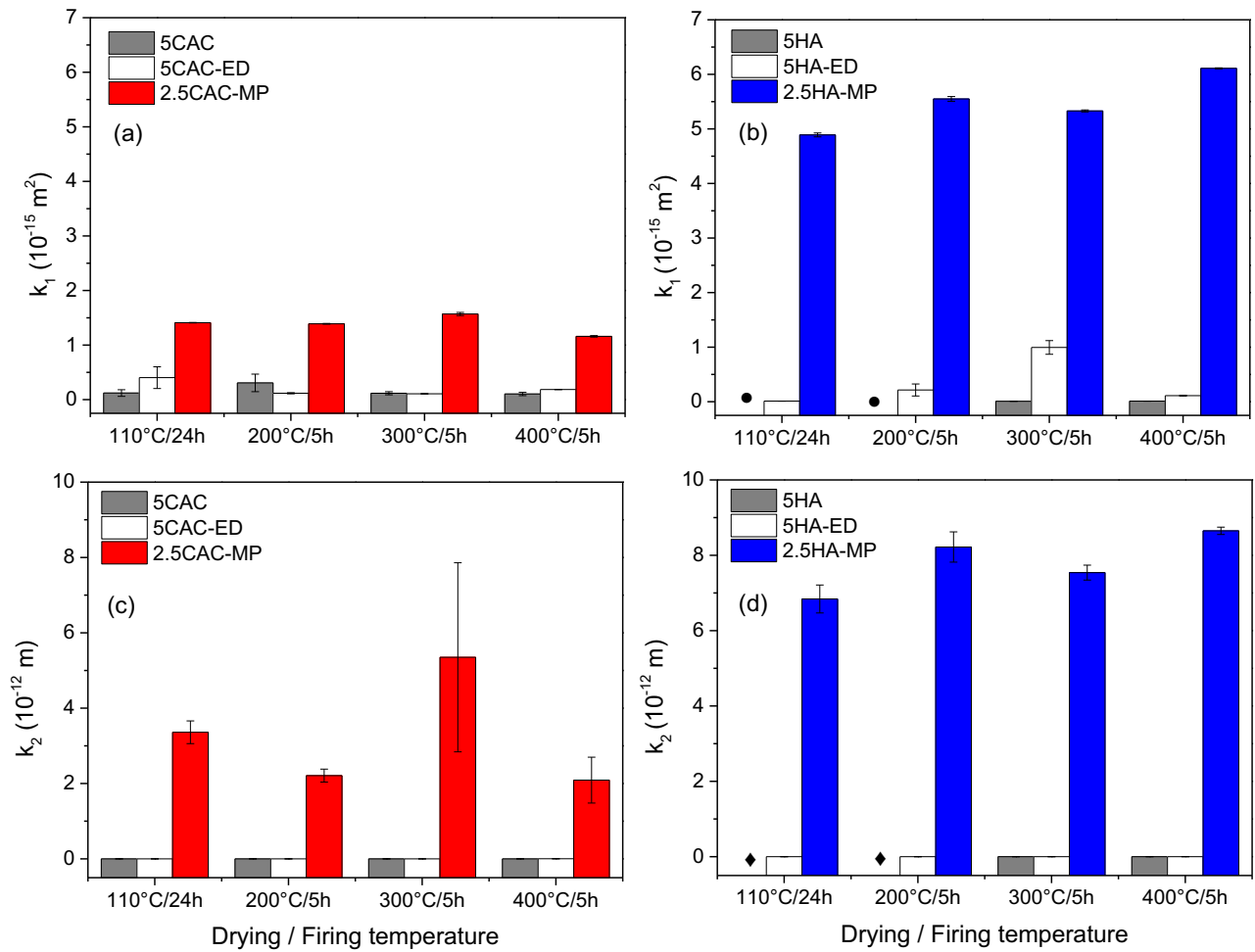

Figure 3 – (a and b) Darcian ($k_1$) and (c and d) non-Darcian ($k_2$) permeability constants of the castable samples obtained after drying (110°C/24h) and firing treatments at different temperatures (200-400°C/5h). ● = $k_1 < 0.9 \times 10^{-17}$ m$^2$ and ♦ = $k_2 < 4.7 \times 10^{-18}$ m for 5HA composition.

Considering that both green mechanical strength and permeability are important parameters that play a role in the drying behavior of dense castables, the following step consisted of evaluating the drying performance and explosion resistance of the cured samples (30°C/24h) during their first thermal treatment. According to Fig. 4a and 4b, no explosion was observed when heating the castables with a 2°C/min rate and the main mass loss took place from 30-120°C for 2.5CAC-MP and 2.5HA-MP compositions, whereas the additive-free and the fiber-containing materials presented more complex profiles with mass changes observed mainly in the following temperature ranges: 30-250°C, 250-350°C and 350-500°C (Fig. 4b). As the evaluated specimens were only cured before testing, the high mass loss identified at the beginning of the TG tests (30-120°C) should be associated to the free-water release and decomposition of gel-like phases. On the other hand, polymeric fibers and crystalline hydrate decomposition (i.e., $C_2AH_8$, $C_3AH_6$, $AH_3$) was mainly detected above 250°C [12-15]. Due to the higher water demand during the preparation of HA-bonded materials, such samples showed increased mass loss when compared to the CAC-based refractories (Fig. 4a).

When increasing the applied heating rates (5 and 20°C/min) during the TG measurements, 5HA, 5HA-ED and 5CAC castables underwent explosion around 200-300°C (Fig. 4c – 4f), as a result of the reduced permeability and high steam pressure generated inside the microstructures. The ED incorporation into the CAC-bonded composition (5CAC-ED) did not induce major changes in the cold flexural strength (Fig. 2a) and permeability constants (Fig. 3a and 3c) of this material when compared to 5CAC. Nevertheless, the effect of the fibers in this system was already enough to inhibit the samples' explosion. The same performance was not achieved for the 5HA-ED composition, which can be explained by both the lower mechanical strength and permeability level of these samples (Fig. 2b, Fig. 3b and 3d). As the 5HA-ED explosion took place at 257°C and the fibers' thermal decomposition should mainly occur above 320°C (Fig. 5) [15], the selected additive was not able to act in its full potential during the TG tests carried out with high heating rates, resulting in a microstructure with a limited number of permeable paths for the water vapor release that, as a consequence, favored the pressurization of the samples. Hence, the polymeric additive must not only melt at low temperatures, but also decompose to generate suitable permeable channels.

The effect of MP addition to 2.5CAC-MP and 2.5HA-MP was characterized by a significant mass loss in the 100-150°C temperature range (Fig. 4a, 4c and 4e), which can be related to the free-water release and decomposition of the gel-like phase derived from the interaction of MP with the binders [12]. These transformations were responsible for increasing the overall permeability of the samples (Fig. 3) and providing better and safer conditions for drying the MP-containing castables. Despite the influence of MP in decreasing the flexural strength of the cured 2.5CAC-MP and 2.5HA-MP samples (Fig. 2), its role in improving the number of permeable paths in the microstructure at the beginning of the heating process (even at 110°C, Fig. 4) is of major importance to inhibit the castables' pressurization.

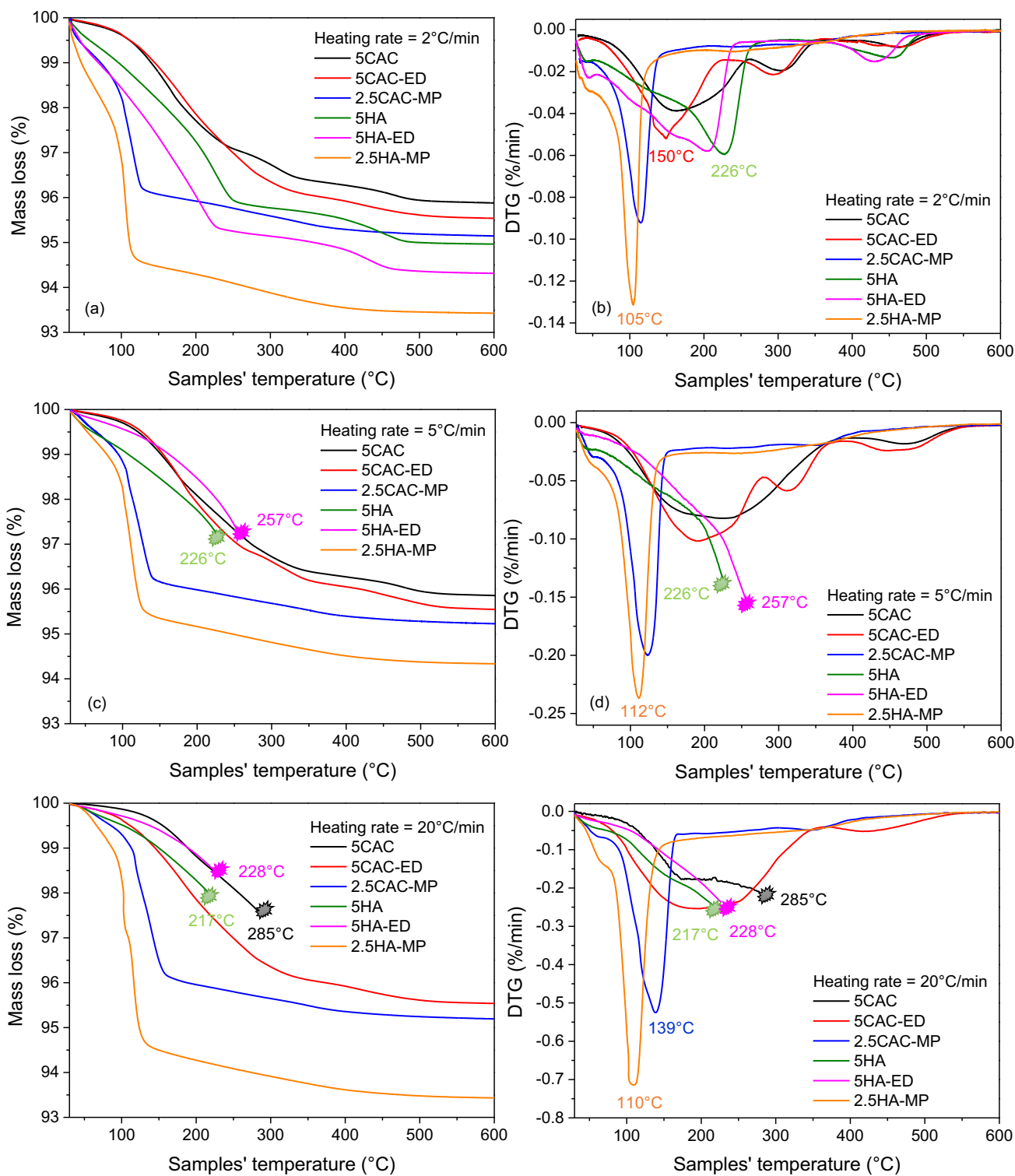

Figure 4 – (a, c and e) Mass loss and (b, d and f) first derivative of mass loss for the evaluated castables. The samples were cured before testing and the measurements were carried out with 2, 5 and 20°C/min heating rate up to 600°C.

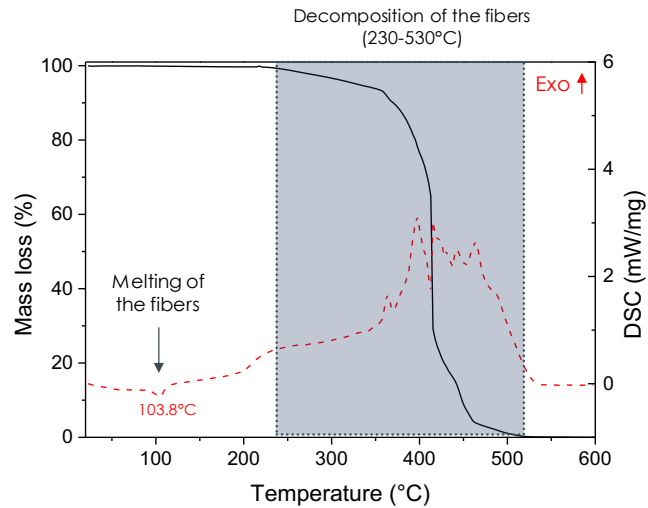

Figure 5 – Mass loss (%) and differential scanning calorimetry (DSC, mW/mg) profiles of the polymeric fibers up to 600°C in oxidizing environment (synthetic air flow = 50 ml/min) and with heating rate = 5°C/min [15].

## 3.2 – Effect of the aluminum salt 2-hydroxypropanoic acid in adjusting the drying behavior of HA-bonded castables

Knowing that 5HA castable is more susceptible to explosion during heating (Fig. 4), due to its reduced permeability and mechanical strength (Fig. 2 and 3), another route (instead of adding ED or MP) should be required to optimize the microstructure of this refractory. Thus, it was decided to study an additional agent to improve the drying behavior of castables containing hydratable alumina.

Fig. 6 indicates that the incorporation of aluminum salt 2-hydroxypropanoic acid (AS) modified the hydration behavior of the used binder, delaying the setting time (velocity increase was shifted to longer times) and reducing the measured velocity values. This behavior is associated with the action of AS, as initially its dissolution should take place in the liquid medium (Eq. 3), changing the overall pH of the refractory system. For instance, after the mixing step, the following pH results were obtained for castables 5HA, 5HA-0.6AS and 5HA-0.6AS-0.5CAC: 9.44, 8.23 and 8.76, respectively.

$$(CH_3CHOHCOO)_3Al_{(s)} \leftrightarrow 3(CH_3CHOHCOO^-)_{(aq)} + Al^{3+}_{(aq)} \qquad (3)$$

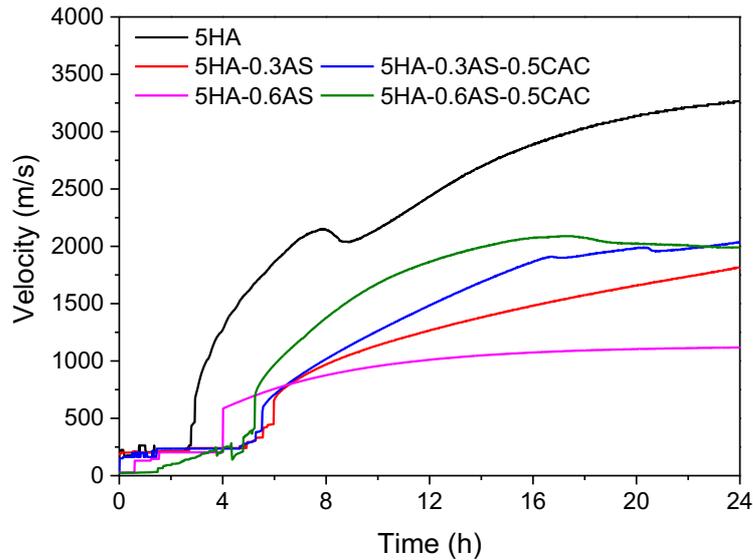

Figure 6 – Curing behavior of the HA-bonded refractory castables obtained via ultrasonic measurements at 22°C for 24h.

The pH decrease usually affects ρ-alumina hydration, changing the overall content of the crystalline (bohemite and bayerite) and amorphous hydrates (gel-like phase) formed in the resulting microstructure [26]. X-ray diffraction of the cured castables' matrix (results not shown here) pointed out that no crystalline compounds were detected in the AS-containing samples, indicating that this additive affected the hydration path of the prepared castables. Besides that, it was not possible to demold the cast samples of these compositions due to their low mechanical strength after curing at 30°C/24h. Hence, a longer curing period (~36h) was required to allow a suitable handling of the 5HA-0.3AS and 5HA-0.6AS specimens. In order to overcome this issue, additional compositions containing AS + 0.5 wt.% of CAC (5HA-0.3AS-0.5CAC and 5HA-0.6AS-0.5CAC) were processed and the collected results highlighted that an improved curing behavior (Fig. 6) and green mechanical strength (Fig. 7a) could be obtained for these castables after a curing time of 24h.

The CAC addition to the AS-containing compositions may firstly induce the generation of a complex compound due to the interaction of the $Ca^{2+}$ ions with the aqueous species resulting from AS dissolution (Eq. 4). Additionally, the dissociation of this complex phase should also favor $Al(OH)_3$ precipitation (Eq. 5) in the resulting mixture, leading to the castable's stiffening in a proper period of time [1].

$$CaO_{(s)} + 2(CH_3CHOHCOO^-)_{(aq)} + H_2O \rightarrow 2(CH_3CHOHCOO^-)Ca^{2+} + 2OH^-_{(aq)} \quad (4)$$

$$Al^{3+}_{(aq)} + 3OH^-_{(aq)} \rightarrow Al(OH)_{3\,(gel)} \quad (5)$$

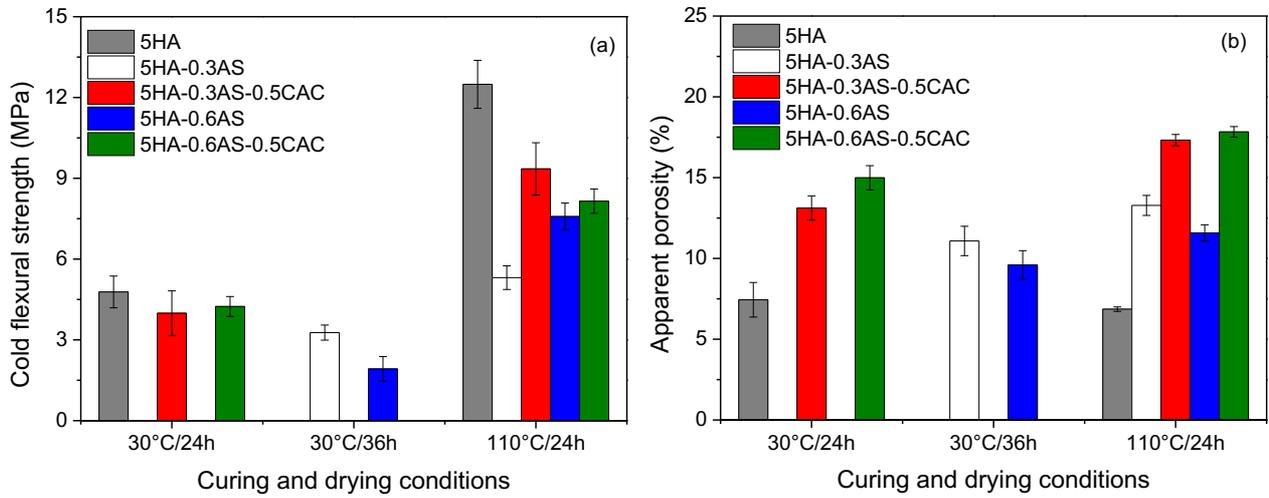

Figure 7 – Cold flexural strength (a) and apparent porosity (b) for the HA-bonded castables after curing (30°C for 24h or 36h) and drying (110°C for 24h).

Cold flexural strength and apparent porosity increase could be observed for the prepared compositions after the drying step at 110°C/24h (Fig. 7a). Despite the low mechanical strength of the AS-containing compositions for the tested conditions, their greater porosity levels might be a positive aspect to prevent thermal crack propagation during a first heating cycle and it could be related to the shrinkage and/or cracking of the gel-like phase generated in the refractories' microstructure [12]. Moreover, when analyzing the permeability constants measured for the dried (110°C) and calcined samples (200-400°C) of the HA-bonded materials, it was observed that AS effectively enhanced $k_1$ and $k_2$ values of the prepared castables, when compared to the results obtained for the reference composition (Tables 3 and 4).

Table 3 – Darcian permeability constant ($k_1$) obtained for the evaluated castables after drying (110°C) and firing (200-400°C).

| $k_1$ (m$^2$) | 5HA | 5HA-0.3AS | 5HA-0.3AS-0.5CAC | 5HA-0.6AS | 5AB-0.6LA-0.5CAC |
|---|---|---|---|---|---|
| 110°C/24h | < 0.90 x 10$^{-17}$ | 6.98 ± 0.07 (10$^{-15}$) | 2.26 ± 0.01 (10$^{-16}$) | 2.45 ± 0.06 (10$^{-16}$) | 6.82 ± 0.08 (10$^{-16}$) |
| 200°C/5h | | 1.84 ± 0.08 (10$^{-14}$) | 2.24 ± 0.04 (10$^{-16}$) | 2.45 ± 0.01 (10$^{-16}$) | 7.02 ± 0.04 (10$^{-16}$) |
| 300°C/5h | 0.90 ± 0.12 (10$^{-17}$) | 188 ± 0.04 (10$^{-13}$) | 2.44 ± 0.03 (10$^{-16}$) | 3.14 ± 0.10 (10$^{-16}$) | 8.32 ± 0.10 (10$^{-16}$) |
| 400°C/5h | 1.13 ± 0.03 (10$^{-17}$) | 3.69 ± 0.20 (10$^{-13}$) | 2.68 ± 0.14 (10$^{-16}$) | 3.42 ± 0.02 (10$^{-16}$) | 8.21 ± 0.05 (10$^{-16}$) |

Table 4 – Non-Darcian permeability constant ($k_2$) obtained for the evaluated castables after drying (110°C) and firing (200-400°C).

| $k_2$ (m) | 5HA | 5HA-0.3AS | 5HA-0.3AS-0.5CAC | 5HA-0.6AS | 5HA-0.6SA-0.5CAC |
|---|---|---|---|---|---|
| 110°C/24h | < 4.70 x 10$^{-18}$ | 2.56 ± 0.04 (10$^{-12}$) | 2.00 ± 0.11 (10$^{-14}$) | 1.06 ± 0.12 (10$^{-14}$) | 3.23 ± 0.21 (10$^{-13}$) |
| 200°C/5h | | 8.94 ± 0.43 (10$^{-12}$) | 2.36 ± 0.62 (10$^{-14}$) | 0.96 ± 0.05 (10$^{-14}$) | 3.48 ± 0.37 (10$^{-13}$) |
| 300°C/5h | 4.70 ± 0.33 (10$^{-18}$) | 2.35 ± 0.02 (10$^{-10}$) | 2.61 ± 0.10 (10$^{-14}$) | 2.18 ± 0.30 (10$^{-14}$) | 3.88 ± 0.48 (10$^{-13}$) |
| 400°C/5h | 1.93 ± 0.12 (10$^{-17}$) | 3.88 ± 0.05 (10$^{-10}$) | 2.62 ± 1.18 ±(10$^{-14}$) | 2.11 ± 0.04 (10$^{-14}$) | 3.21 ± 0.10 (10$^{-13}$) |

In order to evaluate the explosion resistance of the AS-containing castables, thermogravimetric tests of cured samples were carried out up to 600°C using the same heating rates shown in Section 3.1 (2, 5 and 20°C/min). Considering that pseudo-bohemite, bayerite and bohemite decomposition occur in the range of 100-180°C, ~300°C and 450-550°C, respectively [4,6], the main advantage of adding AS to HA-bonded refractories is the likelihood of shifting the most intense mass loss step to earlier temperatures (below 200°C), as indicated in Fig. 8. This performance is related to free-water release and the decomposition of the gel-like complex phase derived from the AS presence in the mix that induced the generation of microstructures with higher porosity and permeable paths (Fig. 7b and Tables 3 and 4), which can prevent the steam pressurization during heating.

As reported in the literature [25], the interaction of CaO (derived from CAC addition to 5HA-0.3AS-0.5CAC and 5HA-0.6AS-0.5CAC) with the selected aluminum salt during the mixing and curing stages might also give rise to $CaC_6H_{10}O_6.5H_2O$, which can decompose approximately at 130°C. This transformation and the withdrawal of free-water may explain the more intense DTG peak of the CAC-containing formulations at temperatures close to 130-150°C (Fig. 8b).

No explosion was observed when conducting the thermogravimetric analyses of the cured castables' samples with a 2°C/min heating rate (Fig. 8a and 8b), which can be related to the longer time available for the withdrawal of the water added to the mixes during their processing steps. However, as expected, a faster drying rate (5°C/min or 20°C/min) induced the explosion of the reference refractory composition (Fig. 8c and 8e). The formulations containing AS and AS+CAC presented a better performance and higher explosion resistance due to the role of such drying additive in increasing the permeability of the resulting microstructure (Tables 3 and 4). When applying a more severe heating condition (20°C/min), only the samples containing higher AS amount and 0.5 wt.% of CAC withstood the thermal and mechanical stresses associated to the drying step (Fig. 8e), indicating that a proper balance between mechanical strength and permeability level of the microstructure could be achieved. Therefore, the selected aluminum salt enabled a large amount of water vapor to be safely released at the initial evaporation stage (100-150°C) of the drying process [9,12], mainly due to the generation of a higher number of permeable paths.

## 4. Conclusions

This work investigated the role of novel commercial drying agents (polyethylene fibers = ED, enhancing a permeability active compound called Refpac Mipore 20 = MP and aluminum salt 2-hydroxipropanoic acid = AS) in dense high-alumina refractory systems containing calcium aluminate cement and hydratable alumina as binders. According to the obtained results, ED was an effective additive to induce a suitable optimization of the CAC-bonded compositions' permeability, preventing their explosion during heating. However, the same positive effect was not achieved when adding these fibers to formulations containing hydratable alumina.

MP was an interesting and versatile alternative route to adjust the castables' microstructure, as its addition to both evaluated systems (formulations bonded with calcium aluminate cement or hydratable alumina) resulted in permeable structures that allowed faster water vapor release in the 100-150°C range. Such behavior was associated with its effect in modifying the hydration reaction sequence of the binders, giving rise to a gel-like phase, instead of crystalline ones, which can undergo decomposition at lower temperatures. Hence, MP was able to provide safer and faster drying procedures for dense castables. The only negative aspect of using this active compound in the designed compositions is the observed decay of the samples' green mechanical strength when compared to the additive-free one.

In the case of using aluminum salt 2-hydroxipropanoic acid in the designed HA-bonded compositions, the action of this additive was similar to the one induced by MP, as it induced changes in the binder hydration reactions. Consequently, the curing behavior and setting time of the AS-containing compositions were delayed and a drop of the green mechanical strength values was observed. Nevertheless, these negative aspects could be minimized when adding 0.5 wt.% of CAC to HA-bonded with AS additive, which resulted in samples with major mass loss at relatively low temperatures (< 200°C) during heating and suitable mechanical strength and permeability values. Castable 5AB-0.6AS-0.5CAC was the most promising compositions as it withstood even the most severe tested condition (heating rate = 20°C/min) without presenting explosion of the samples.

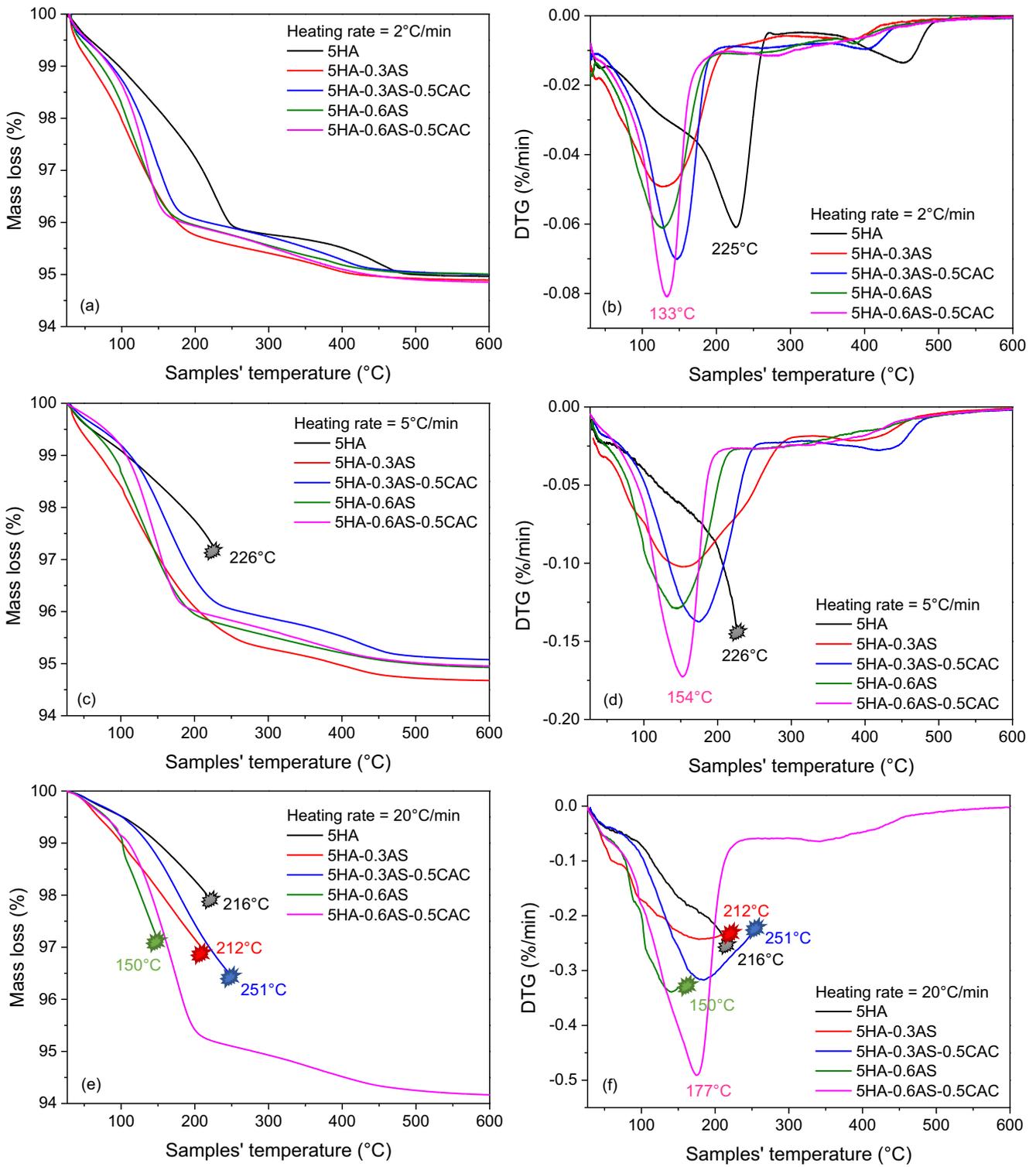

Figure 8 – (a, c and e) Mass loss profiles and (b, d and f) first derivative of the mass loss results for the HA-bonded refractory castables.

## 5. Acknowledgements

This study was financed in part by the Coordenação de Aperfeiçoamento de Pessoal de Nível Superior - Brasil (CAPES) - Finance Code 001 and grant number: 303324/2019-8. The authors would like to thank Fundação de Amparo a Pesquisa do Estado de São Paulo (FAPESP, grant number 2019/07996-0) for supporting this work and Almatis (Brazil), Elkem (Norway) and Imerys Aluminates S.A. (France) for supplying the raw materials used in this work.